\documentstyle[12pt, aps, preprint, eqsecnum]{revtex}
\begin{document}

\title{ PATH INTEGRAL SOLUTION OF NON-CENTRAL POTENTIAL
 }

\vspace{.4in}

\author{Bhabani Prasad Mandal}

\address{
Theory Group, Saha Institute of Nuclear Physics, \\
1/AF Bidhannagar, Calcutta-700064, India.\\
\mbox{email: \ bpm@tnp.saha.ernet.in } \\}

\vspace{.4in}

\maketitle

\vspace{.4in}

\begin{abstract}
We have studied the path integral solution of a system of particle moving in
certain class of non-central potential without using Kustannheimo-Stiefel
transformation. The Hamiltonian of the system has been converted
to a separable Hamiltonian of Liouville type in parabolic coordinates and
has further reduced to a Hamiltonian corresponding to two 2- dimensional
simple harmonic oscillators.
The energy spectrum for this system is calculated analytically.
 Hartmann ring-shaped potential and compound
Coulomb plus Aharanov- Bohm potential have also been studied  as special
cases.
\end{abstract}
\newpage
\section{ Introduction}

The path integral approach to quantum mechanical systems is elegant in
formalism and powerful in semi-classical calculations \cite{fey}. However, for many of
the quantum mechanical systems, it is very difficult to carry out explicit
path integrals. In 1979, Duru and Kleinert \cite{duru} developed some method using
the so called Kustaanheimo -Stiefel (KS) transformation \cite{ks} 
to study the path integral of hydrogen atom problem. Many of the
clarifying works have been published in this approach in last 15 years 
\cite{ter,gro,gan,che,ho,wood,ste,vic,ast,ast1} .
Particularly, Nersessian and Ter-antonian \cite{ter} have studied the system of
dyon using this  KS transformation. Carpio  et al. \cite{vic} have obtained energy
dependent Green's functions for a particle in Hartmann potential \cite{hat}. Path integral
solution of a non-isotropic Coulomb like potential has been studied by
Grosche \cite{gro}. Chetouani and et al. \cite{che} have given the algebraic treatment of the Kaluza-
Klein mono pole system. This KS transformation has also been used in many other
fields of physics \cite{ast,ast1}. In  all the above works, the complicated
Hamiltonians
have been reduced to relatively simple Hamiltonians in higher dimension by the
use
of KS transformation.

However, the use of KS transformation is highly  technical and it is not
necessary to solve many of the above problems. Further, it has it's
own complications. 
Recently Fujikawa \cite{fu} has analyzed the path integral of hydrogen atom problem by
regarding it as a special case of the general treatment of classically
separable Hamiltonian of Liouville type and thus has avoided the use of
the KS transformation . The basic dynamical principle involved in his 
formulations  is Jacobi's principle of least action for a given energy. The
feature of the path integral on the basis of Jacobi's principle of least
action is that it is basically static and it is reparametrization
invariant. One particular application has been discussed in Ref. [11] 
where the Hamiltonian  corresponding to hydrogen atom  in electric
field has been shown to be reduced to a Hamiltonian corresponding to
 anharmonic oscillator with an unstable quartic coupling. 
This approach may find many other applications in future.

In this work we have extended Fujikawa's work for the case of 
non-central potential of the kind,

\begin{equation}
V(r, \theta ) = -\frac{ Ze^2}{r} + \frac{ B\hbar^2}{2mr^2\sin^2 \theta }
+ \frac{ C\hbar^2\cos\theta }{2mr^2\sin^2 \theta}
\label{pot}
\end{equation}
where $B $ and $C$ are real constants.  Instead of using KS transformation to study  this problem, 
 we have followed the approach of Ref. [11] 
and have converted the  Hamiltonian of non-central potential
into a separable Hamiltonian of Liouville type in parabolic coordinate system.
In this parabolic coordinate system, using some canonical transformation,
we have shown that this Hamiltonian can be further reduced to a Hamiltonian corresponding
to  two 2-dimensional simple harmonic oscillators. Hence we evaluate the 
path integral for the particle in a non-central potential exactly and
obtain the Green's functions for the system. We
further derived the energy spectrum of this system analytically in a very simple fashion.
The importance of the potential in Eq. \ref{pot} lies on the fact that
compound Coulomb plus Aharanov-Bohm potential \cite{ab}  and Hartmann ring-shaped 
potential, originally proposed as model for the Benzene molecule \cite{hat}, are mathematically 
linked to this
potential. In fact the energy spectrum for these two 
potentials can be obtained directly by considering these as a special case of
the general non-central potential.

In the next section, the parabolic coordinates are used to simplify
our Hamiltonian to convert it into  a separable Hamiltonian of Liouville type.
We have obtained the path integral for this non-central potential exactly
using this separable Hamiltonian. In section III,  we have defined our
Hamiltonian in terms of creation and annihilation operators corresponding
to simple harmonic oscillators. Thus we have obtained analytically the bound state energy
spectrum for this system in a very simple way. In the following section, we have
considered
Hartmann ring shaped potential \cite{hat} and Aharanov-Bohm potential \cite{ab}  as special
cases of the general non-central potential. Energy spectrums for both the
cases have been obtained. Finally we conclude by summarizing our results.

\section{  Path Integral of Non- Central Potential}
The most general Hamiltonian of Liouville type can be written as
\begin{equation}
H= \frac{1}{ V_1(q_1) +V_2(q_2)}\left\{ \frac{1}{2mW_1(q_1)} p_1^2 + \frac{1}{
2mW_2(q_2)}p^2_2 + U_1(q_1) +U_2(q_2) \right\} 
\label{21}
\end{equation}
This can be reduced to a simpler form by using some canonical
transformation
and redefining $U$ 's and $V$'s
[See Ref. [11]] like
\begin{equation}
H= \frac{1}{ V_1(q_1) +V_2(q_2)}\left\{ \frac{1}{2m} p_1^2 + \frac{1}{
2m}p^2_2 + U_1(q_1) +U_2(q_2) \right\}
\label{22}
\end{equation}

The Schroedinger Eq. $H\phi = E\phi $ can be rewritten as $\hat{H}_T\phi
=0$,
where $\hat{H}_T (= H-E )$ is total Hamiltonian.

\begin{equation}
\hat{H}_T = \frac{ 1}{2m} \left ( \hat{p}_1^2 +\hat{p}_1^2 \right ) +
U_1(q_1) +U_2(q_2) - E \left ( V_1(q_1) +V_2(q_2) \right )
\label{100}
\end{equation}
Now, to obtain the path integral for this Hamiltonian $\hat{H}_T$, let us  consider
the evaluation of this operator for a arbitrary parameter $\tau$ :
\begin{eqnarray}
\left <q_{1b},\, q_{2b} \left |\exp \left \{-i  \frac{ \hat{H}_T\tau}{\hbar}\right \}
\right | q_{1a}, \, q_{2a} \right >= 
\left < q_{1b} \left | \exp \left \{ -( \frac{ i}{\hbar})\left [ \frac{ 1}{2m}\hat{p}_1^2 +U_1(q_1) -EV_1(q_1)
\right ] \tau \right \} \right | q_{1a} \right > \nonumber \\
\times   \left < q_{2b} \left | \exp \left \{ -( \frac{ i}{\hbar})\left [ \frac{ 1}{2m}\hat{p}_2^2 +U_2(q_2)
-EV_2(q_2)
\right ] \tau \right \} \right | q_{2a} \right >
\label{200}
\end{eqnarray}
The RHS of Eq. \ref{200} can be written in terms of path integral
\cite{fey}.
\begin{eqnarray}
\left <q_{1b},\, q_{2b} \left |\exp \left \{-i  \frac{ \hat{H}_T\tau}{\hbar}\right \}
\right |q_{1a}, \, q_{2a} \right >\mbox{\hspace{3in}} \nonumber \\= 
 \int {\cal D }q_1{\cal D}p_1 \exp \left \{ ( \frac{ i}{\hbar}\int _{0}^{\tau}
\left [ p_1\dot{q_1} -\left ( \frac{ \hat{p}_1^2}{2m}+U_1(q_1) -EV_1(q_1) \right )
\right ] d\tau \right \}\nonumber \\
\times\int {\cal D }q_2{\cal D}p_2\exp \left \{ ( \frac{ i}{\hbar}\int _{0}^{\tau}
\left [ p_2\dot{q_2} -\left ( \frac{ \hat{p}_2^2}{2m}+U_2(q_2) -EV_2(q_2) \right )
\right ] d\tau \right \}
\label{201}
\end{eqnarray}
The parameter $\tau $ is arbitrary and one can obtain physically meaningful
quantity out of Eq. \ref{200} by integrating  over $\tau $ from $0$ to
$\infty$ \cite{fu}.
\begin{eqnarray}
\left <q_{1b},\, q_{2b} \left |   \frac{\hbar}{ \hat{H}_T}
\right | q_{1a}, \, q_{2a} \right >_{\mbox{Semi-classical}}\mbox{\hspace{3.6in}}\nonumber \\
=\int _0^\infty d\tau
\left < q_{1b} \left | \exp \left \{ -( \frac{ i}{\hbar})\left [ \frac{ 1}{2m}\hat{p}_1^2 +U_1(q_1) -EV_1(q_1)
\right ] \tau \right \} \right | q_{1a} \right > \nonumber \\
\times   \left < q_{2b} \left | \exp \left \{ -( \frac{ i}{\hbar})\left [ \frac{ 1}{2m}\hat{p}_2^2 +U_2(q_2)
-EV_2(q_2)
\right ] \tau \right \} \right | q_{2a} \right >
 \label{2002}
\end{eqnarray}
The meaning of RHS of Eq. \ref{2002} has been explained in Ref. [11].
And the LHS of it can be written as \cite{fu}, 
\begin{equation}
\left <q_{1b},\, q_{2b} \left |   \frac{\hbar}{ \hat{H}_T}
\right | q_{1a}, \, q_{2a} \right > =
\left <q_{1b},\, q_{2b} \left |   \frac{\hbar}{ \hat{H}-E}
\right | q_{1a}, \, q_{2a} \right > \frac{1}{V_1(q_{1a}) +V_2(q_{2a})}
\label{202}
\end{equation}
with the completeness relation
\begin{equation}
\int\left | q_1\  q_2 \right >\frac{dV}{V_1(q_1)+V_2(q_2)}\left < q_1 \ q_2 \right | =1
\label{203}
\end{equation}

The LHS of Eq. \ref{202} thus defines the correct Green's functions for the operator
$ \frac{1}{\hat{H} -E}$ .

In this section, we will  show how the Hamiltonian  of a particle in a
non-central potential can be reduced to a separable Hamiltonian of the
above
kind.
Let us start with the non-central potential  written in Eq. \ref{pot}
\begin{equation}
V(r, \theta ) = -\frac{ a}{r} + \frac{ b}{r^2\sin^2 \theta }
+ \frac{ c\cos\theta }{r^2\sin^2 \theta}
\label{23}
\end{equation}
where 
\begin{equation}
 a = Ze^2 ;\ \ \ \ \ \ b = \frac{B\hbar^2}{2m}; \ \ \ \ \ \ c = \frac{C\hbar^2}{2m}
\end{equation}
The coulomb and the ring- shaped potentials are particular case  of this potential. 
To express the potential in parabolic coordinates $(\xi, \eta, \phi)$,
it is useful to first  express this potential $ V(r,\theta)$ in cylindrical coordinates
$(\rho, \phi, z)$. 
 In cylindrical coordinate the potential looks like,
\begin{equation}
V(\rho,z) = -\frac{a}{\sqrt{\rho^2 +z^2}} + \frac{b}{\rho^2} +\frac{cz}{\rho^2
\sqrt{\rho^2+z^2}}
\label{25}
\end{equation}
The parabolic coordinates are defined in terms of cylindrical coordinates as
\begin{eqnarray}
\xi &=& \frac{1}{2}\left ( \sqrt{\rho^2+z^2} -z \right ) \nonumber\\
\eta &=&\frac{1}{2}\left (\sqrt{\rho^2+z^2} + z \right ); \ \ \ \  
\phi = \phi \label{26}
\end{eqnarray}
Now the potential in Eq. \ref{25} in terms of these parabolic coordinates,
is
\begin{equation}
V(\xi,\eta) = -\frac{a}{\xi+\eta}+ \frac{ b}{4\xi\eta} + \frac{
c(\eta-\xi)}{4\eta\xi(\eta+\xi)}\label{27}
\end{equation}
Therefore the Hamiltonian in parabolic coordinate
can be written as
\begin{equation}
H(\xi,\eta,\phi) = \frac{ 1}{2m(\xi+\eta)}\left [\xi \hat{p}_{\xi}^2 +
\eta \hat{p}_\eta^2 \right ] +\frac{ 1}{8m\eta\xi} \hat{p}_\phi^2 +V(\xi,\eta)
\label{28}
\end{equation}
Let us  further perform a canonical transformation which simplifies
the kinetic term in $H$ in Eq. \ref{28}.
\begin{eqnarray}
\xi &=& \frac{ 1}{4} u^2 ;\ \ \ \ \ 0\leq u<\infty \nonumber\\
\eta &=&\frac{ 1}{4} v^2 ;\ \ \ \ \ 0\leq v<\infty \label{29}
\end{eqnarray}
This implies the relation between the canonical conjugate momentum
variables as
\begin{eqnarray}
\sqrt{\xi}\hat{p}_\xi &=&\hat{ p}_u \nonumber\\
\sqrt{\eta}\hat{p}_\eta &=& \hat{p}_v \label{210}
\end{eqnarray}
Using Eqs. \ref{29} and \ref{210} in the expression for Hamiltonian
in \ref{28} we arrived at
\begin{equation}
H(u,v,\phi) = \frac{ 4}{u+v}\left \{ \frac{ 1}{2m} \left [\hat{p}_u^2 +\hat{p}_v^2
+(\frac{ 1}{ u^2}+ \frac{ 1}{v^2})\hat{p}_\phi^2 \right ] -a + \frac{ b+c}{u^2}
+ \frac{ b-c}{v^2} \right \} \label{211}
\end{equation}
This can be written compactly as 
\begin{equation}
H(u,v,\phi) = \frac{ 4}{u^2+v^2} \left \{ \frac{ 1}{2m} \left [\hat{ p}_u^2
+\hat{p}_v^2 +\frac{ 1}{u^2}\hat{p}_{\phi_1}^2 +\frac{
1}{v^2}\hat{p}_{\phi_2}^2 \right ] -a \right \}
\label{212}
\end{equation}
where
\begin{equation}
\hat{p}_{\phi_1}^2 = \hat{p}_{\phi}^2 + 2m(b+c);\;\;\;\; 
\hat{p}_{\phi_2}^2 = \hat{p}_{\phi}^2 + 2m(b-c) 
\label{213}
\end{equation}
This Hamiltonian is still not a separable one of Liouville type.
We will further consider the total Hamiltonian $H_T$.

\begin{eqnarray}
\hat{H}_T &=& \frac{ 1}{2m} \left [\hat{p}_u^2 
+\hat{p}_v^2 +\frac{ 1}{u^2}\hat{p}_{\phi_1}^2 +\frac{
1}{v^2}\hat{p}_{\phi_2}^2 \right ] -a- \frac{ E(u^2+v^2)}{4} \nonumber \\
 &=& \frac{ 1}{2m} \left [\hat{p}_u^2 
+\hat{p}_v^2 +\frac{ 1}{u^2}\hat{p}_{\phi_1}^2 +\frac{
1}{v^2}\hat{p}_{\phi_2}^2 \right ] -a+ \frac{1}{2}m\omega^2 (u^2+v^2) 
\label{214}
\end{eqnarray}
where we have considered the bound state case ($E< 0$) and $\omega $ is
defined as 
\begin{equation}
\omega^2 = -\frac{ E}{2m}
\label{ww}
\end{equation}
Now we introduced 2-dimensional vectors $\vec{u} $ and $\vec{v}$ as
\begin{eqnarray}
\vec{u} &=& ( u_1, u_2) = \left ( u\cos\phi_1 , u\sin\phi_1 \right )
\nonumber \\
\vec{v} &=& ( v_1, v_2) = \left ( v\cos\phi_2 , v\sin\phi_2 \right )
\label{215}
\end{eqnarray}

Then we have
\begin{equation}
\vec{p_u}^2 = \hat{p}_u^2 + \frac{ 1}{u^2}\hat{p}_{\phi_1}^2;
\;\;\;\;\;\;\;
\vec{p_v}^2 = \hat{p}_v^2 + \frac{ 1}{v^2}\hat{p}_{\phi_2}^2;\label{216}
\end{equation}
Putting all these in 
Eq. \ref{214}; 
\begin{equation}
\hat{H}_T = \frac{ 1}{2m} \vec{p}_u^2 +\frac{ m\omega^2}{2}\vec{u}^2 +
\frac{ 1}{2m} \vec{p}_v^2 + \frac{ m\omega^2}{2}\vec{v}^2 -a
\label{217}
\end{equation}
This is the Hamiltonian which is separable of Liouville type. Now
by using Eq. \ref{200} for this separable Hamiltonian, we can find the path integral 
for the non central potential exactly as
\begin{eqnarray}
\left < \vec{u}_b ,\, \vec{v}_b \left |\exp \left [-i \frac{
\hat{H}_T\tau}{\hbar} \right ] \right |\vec{u}_a ,\, \vec{v}_a \right > =
e^{\frac{ ia\tau}{\hbar}}
 \left < \vec{u}_b \left |\exp \left [ - (\frac{
i}{\hbar})\left ( \frac{ \vec{p}_u^2}{2m} + \frac{ m\omega^2}{2}\vec{u}^2 \right )\tau \right ] \right
| \vec{u}_a \right >\nonumber \\
\times \left < \vec{v}_b \left |\exp \left [ - (\frac{
i}{\hbar})\left ( \frac{ \vec{p}_v^2}{2m} + \frac{ m\omega^2}{2}\vec{v}^2 \right )\tau \right ] \right
| \vec{u}_a \right >\nonumber \\
=e^{(\frac{ ia\tau}{\hbar})}{ \left (\frac{i m\omega}{2\pi i\hbar\sin\omega\tau}\right )}^{
\frac{ 4}{2}}\exp \left \{ \frac{ im\omega}{2\hbar\sin\omega\tau} \left [
\left ( \vec{u}_b^2 + \vec{v}_b^2 +\vec{u}_a^2 +\vec{v}_a^2 \right
)\cos\omega\tau - 2\vec{u}_b\vec{u}_a -2\vec{v}_b\vec{v}_a \right ] \right
\}\label{pt}
\end{eqnarray}
where  the exact result  for one dimensional simple harmonic oscillator has been used \cite{fey}.
\begin{equation}
\left < q_b \left | \exp \left [ - \frac{ i}{\hbar} \left ( \frac{
\hat{p}^2}{2m}
+ \frac{ m\omega^2 q^2}{2} \right  )\tau \right ] \right | q_a \right > =
\left ( \frac{ m\omega}{2\pi i \hbar\sin \omega\tau} \right )^{ \frac{ 1}{2}}
\exp \left \{ \frac{ im\omega}{2\hbar\sin \omega\tau} \left [ (q_a^2 +q_b^2)
\cos \omega\tau - 2q_aq_b \right ] \right \}
\end{equation}

The Eq. \ref{pt} 
 contains the arbitrary parameter
$\tau$ and has to be eliminated to obtain physically meaningful quantity.
This can be done by integrating over $\tau$ from $0$ to $\infty$ in both side
of the Eq. \ref{pt}. When we integrate over $\tau$ in the LHS of the
Eq. \ref{pt},
it is nothing but  the Green's functions of the operator $\frac{1}{\hat{H} -E}$ as discussed 
at beginning of this section.
And the integration in the RHS can be done in a  straightforward manner \cite{fu}.
Thus we obtain the explicit Green's functions  for the system of non central
potential.

\section{ Energy Levels}

In the previous section, we have seen that the Hamiltonian for the non-central
potential has been reduced to that of two 2-dimensional oscillators (apart from some
constant shift in ground state energy )[see Eq. \ref{217}].  Therefore we can obtain the energy levels
for this system of non central potential in a simple algebraic way.
Let us define  the creation and annihilation operators for this theory as
 \begin{eqnarray}
a_k = \frac{ 1}{\sqrt{2}} \left [ \sqrt{ \frac{ m\omega}{\hbar}}u_k + \frac{ i}{
\sqrt{m\omega \hbar}}\hat{p}_{u_k} \right ] \nonumber \\ 
\tilde{a}_k = \frac{ 1}{\sqrt{2}} \left [ \sqrt{ \frac{ m\omega}{\hbar}}v_k + \frac{ i}{
\sqrt{m\omega \hbar}}\hat{p}_{v_k} \right ]  
\label{31}
\end{eqnarray}
where $k= 1, 2$ .
The total Hamiltonian in Eq. \ref{217} can be written in terms of these creation
and annihilation  operators as follows:
\begin{equation}
\hat{H}_T = \hbar\omega \left [ \sum _{k=1} ^ 2 \left (a_k^{\dagger}a_k +
\tilde{a}_k^{\dagger}a_k \right ) +2 \right ] -a
\label{32}
\end{equation}
And the conjugate momentum variables can be written as
\begin{eqnarray}
\hat{p}_{\phi_1} = i\hbar \left [ a_1^\dagger a_2 - a_2^\dagger a_1 \right
]\nonumber \\
\hat{p}_{\phi_2} = i\hbar \left [ \tilde{a}_1^\dagger \tilde{a}_2 -
\tilde{a}_2^\dagger\tilde{
a}_1
\right ]\label{33}
\end{eqnarray}
Now, we want to use an unitary transformation of the following type,
\begin{eqnarray}
a_1 = \frac{ 1}{\sqrt{2}} \left ( b_1 - ib_2 \right )\nonumber \\
a_2 = \frac{ 1}{\sqrt{2}} \left ( -ib_1 + b_2 \right )
\label{34}
\end{eqnarray}
and similar transformations for $\tilde{a}_1 , \tilde{a}_2$ also.
Using these transformations in Eqs. \ref{32} and \ref{33} we get,
\begin{equation}
\hat{H}_T = \hbar\omega \left [\sum_{k=1}^2 \left (b_k^\dagger b_k +
\tilde{b}_k^\dagger \tilde{b}_k \right )+2 \right ] -a \
\label{35}
\end{equation}
and 
\begin{eqnarray}
\hat{p}_{\phi_1} = \hbar \left [b_1^\dagger b_1 - b_2^\dagger b_2 \right ]
\nonumber \\
\hat{p}_{\phi_2} = \hbar \left [\tilde{b}_1^\dagger\tilde{ b}_1 - \tilde{b}_2^\dagger\tilde{ b}_2 \right ]
\label{36}
\end{eqnarray}
Now we define the number operators
\begin{eqnarray}
n_k = b_k^\dagger b_k \nonumber \\
\tilde{n}_k = \tilde{b}_k^\dagger \tilde{b}_k
\label{37}
\end{eqnarray}
In terms of number operators the total Hamiltonian in Eq. \ref{35} now can be written as
\begin{eqnarray}
\hat{H}_T &=& \hbar\omega \left [ n_1 + n_2 + \tilde{n}_1 +\tilde{n}_2 +2
\right ] -a \nonumber \\
&=& \hbar\omega \left [ 2n_2 +2\tilde{n}_2 +\frac{ \hat{p}_{\phi_1}}{\hbar}
+ \frac{ \hat{p}_{\phi_2}}{\hbar} +2 \right ] -a
\nonumber \\
&=& 2\hbar\omega \left [n_2 +\tilde{n}_2+1+ \frac{ \hat{p}_{\phi_1} +\hat{p}_{\phi_2}}{\hbar}
\right ] -a
\end{eqnarray}
Here we have used the relation  of Eq. \ref{36}.

The physical state condition is
\begin{equation}
\left [ 2(n_2+\tilde{n}_2 +1) \hbar\omega + \omega \left ( \hat{p}_{\phi_1} +\hat{p}_{\phi_2}
\right ) -a \right ] \phi_{phy} = 0 \label{kk}
\end{equation}
Now let's say,
\begin{equation}
\left [\hat{p}_{\phi_1} + \hat{p}_{\phi_2} \right ] \phi_{phy}\equiv
\lambda \phi_{phy}\label{jj}
\end{equation}
Then Eq. \ref{kk} can  be written as
\begin{equation}
\left [ 2(n_2+\tilde{n}_2+1) \hbar\omega + \omega \lambda  
-a \right ] \phi_{phy} = 0 \label{kk1}
\end{equation} 
i.e.
\begin{equation} 
\omega^2 = \frac{ a^2}{\left [2(n_2+\tilde{n}_2 +1)\hbar + \lambda  \right ]^2}
\label{kk2}
\end{equation}
Hence the energy level can be written in terms of $\lambda $ using the Eq. \ref{ww}
as
\begin{equation}
E_{n_2,\tilde{n}_2,\lambda } = -\frac{ 2ma^2}{\left [2(n_2+\tilde{n}_2+1)\hbar +\lambda \right ]^2}
\label{kk3}
\end{equation}
Using Eqs. \ref{jj} and \ref{213}  $,\lambda
$ can be calculated easily
as 
\begin{equation} 
\lambda  = \hbar \left [ \left (\nu^2 +B+C \right )^{ \frac{ 1}{2}} +
\left (\nu^2+B-C \right )^{ \frac{ 1}{2}} \right ] 
\end{equation}
where $\nu$ is non-negative integer.
Therefore, the complete bound state spectrum for the problem is
\begin{equation}
E_{n_2,\tilde{n}_2,\nu} =  \frac{- m Z^2e^4}{2\hbar^2 \left [
n_2+\tilde{n}_2 +1 +\frac{\sqrt{\nu^2 +B+C} + \sqrt{\nu^2 +B-C}}{2} \right ]^2}
\label{ene}
\end{equation}

This result agrees with that of in Refs. [7,14] where energy spectrum has 
been calculated by solving Schroedinger equation using KS transformation.

\section{ Special Cases}
\begin{itemize}
\item {\bf A. Hartmann ring-shaped potential}

Hartmann's ring-shaped potential is given by
\begin{equation}
V(r, \theta ) = \gamma \sigma^2 \left ( \frac{ 2a}{r} - \frac{ \gamma
a^2}{r^2\sin^2 \theta } \right ) E_0
\label{hp}
\end{equation}
where $a= \frac{ \hbar^2}{me^2}$ and $E_0 = -\frac{ me^4}{2\hbar^2}$,
are Bohr's radius and energy of the ground state of the
hydrogen atom respectively; $\gamma $ and $\sigma$ are dimensionless positive
parameters. This potential had been proposed by Hartmann as a model 
for the Benzene molecule \cite{hat}. 
This potential can be obtained from the non-central potential 
of Eq. \ref{pot} by setting $ C=0, B= \gamma ^2 \sigma^2$ and $Z=\gamma
\sigma^2$.
Therefore, the energy spectrum this system can written immediately
from Eq. \ref{ene} as
\begin{equation}
E_{n_2,\tilde{n}_2,\nu} =  \frac{- m \gamma ^2\sigma^2 e^4}{2\hbar^2 \left [
n_2+\tilde{n}_2 +1 +\sqrt{ \nu^2+\gamma ^2\sigma^2}  \right ]^2}
\end{equation}

\item {\bf B.  Aharanov- Bohm potential:} \\
The Aharanov- Bohm  vector potential, $A$ can be defined in spherical
coordinates ( $ r ,\theta ,\phi$) as
\begin{equation}
A_r = A_{\theta } =0 \ \ \ \mbox{and} \ \  A_{\phi} = \frac{ F}{2\pi r \sin \theta }
\label{abp}
\end{equation}
where $F$ = constant flux created within a very thin and infinitely  long
solenoid, oriented along the $Z$ axis. A punctual charge ($-e$)  is located
at the origin of the axes. The Hamiltonian of the composite system
is given by 
\begin{equation}
H= -\frac{ \hbar^2}{2m}\Delta^2 - \frac{ Ze^2}{r} + \frac{ 1}{2m} \left [
\frac{ A_0}{r^2\sin^2 \theta }+ \frac{ iB_0}{r^2\sin^2 \theta } \frac{ \partial }{
\partial \phi} \right ]  
\label{abh}
\end{equation}
with
$A_0= \left ( \frac{ ZeF}{2\pi c} \right )^2 $ and $ B_0 = \frac{ Ze\hbar
F}{\pi c}$. 
By considering the $\phi$ dependence of the wave function as
$e^{i\nu \phi}$, the effective potential for this system turn out
to be 
\begin{equation}
V = -\frac{ Ze^2}{r} + \frac{ 1}{2m} \left [ \frac{ A_0}{r^2\sin^2 \theta }
-\frac{ B_0\nu}{r^2\sin^2 \theta } \right ]
\label{abe}
\end{equation}
which is a special case of our non-central potential of Eq. \ref{pot}
with $ c=0,\ \ b= \frac{ 1}{2m} (A_0 -B_0\nu)$. 
Now we calculate the value of $\lambda $ using Eqs. \ref{jj} and \ref{213}
for the special situation and put in Eq. \ref{ene} to obtain the spectrum
for the Coulomb plus Aharanov- Bohm system.
\begin{equation}
E_{n_2,\tilde{n}_2,\nu} =  \frac{- m Z^2e^4}{2\hbar^2 \left [
n_2+\tilde{n}_2 +1 +|M| \right ]^2}
\end{equation}
where $|M| =  |\nu - \frac{ ZeF}{2\pi \hbar c}|$.

 If the flux is quantized  
\begin{eqnarray} 
F& =& \frac{ 2\pi\hbar c}{Ze} [\nu -|M|]
\nonumber \\
 &=& \frac{ 2\pi\hbar c}{Ze}\times \mbox{integer}
\end{eqnarray}
then $|M|$ is integer. And 
when $|M|$ is integer, the spectrum thus obtained is nothing but a purely Coulombian 
spectrum.
This  means there is  no Aharanov- Bohm effect.

\end{itemize}

\section{ conclusion}
We have studied the system of particle moving in certain class of
non- central potential without using complicated Kustannheimo-Stiefel
transformation. The complicated Hamiltonian corresponding to the potential
in Eq. \ref{pot} has been reduced to a Hamiltonian of two 2D harmonic
oscillators in parabolic coordinates using some simple trick, mentioned
in Ref. [11]. Next we have calculated the Green's functions for the
system using path integral method for this separable Hamiltonian of
Liouville type.
We have written this Hamiltonian of two 2D harmonic oscillator in terms of
creation and annihilation operators and have obtained the full spectrum
of the system in a simple analytic way. Further, we have studied
Aharanov-Bohm potential and ring-shaped Hartmann potential as
the special cases of the non-central potential and obtained
energy spectrum for these cases. This method of solving quantum
mechanical problems may be useful in solving other complicated
systems analytically.

\begin{center}
{\bf Acknowledgments}
\end{center}

This work is done with partial support of ICSC - World Laboratory
(Lausanne) fellowship.

\end{document}